\documentclass[aps,prb,amsmath,amssymb,twocolumn,10pt]{revtex4-1}

\usepackage{graphicx}
\usepackage{bm}% bold math
\usepackage{makeidx}
\usepackage{hyperref}

\begin{document}
\title{Frustrated Magnetism and Caloric Effects in Mn-antiperovskite Nitrides: $Ab~Initio$ Theory}
\author{J.~Zemen,$^{1,2}$ E.~Mendive-Tapia,$^{3}$ Z.~Gercsi,$^{1,4}$ R.~Banerjee,$^{5}$ J.B.~Staunton,$^{3}$ and K.G.~Sandeman$^{1,6,7}$}
\affiliation{$^{1}$Department of Physics, Blackett Laboratory, Imperial College London, London SW7 2AZ, United Kingdom}
%\affiliation{$^{2}$Institute of Physics ASCR, v. v. i., Cukrovarnick\'a 10, 162 00 Praha 6, Czech Republic}
\affiliation{$^{2}$Faculty of Electrical Engineering, Czech Technical University in Prague, Technick\'a 2, Prague 166 27, Czech Republic}
\affiliation{$^{3}$Department of Physics, University of Warwick, Coventry CV4 7AL, United Kingdom}
\affiliation{$^{4}$CRANN and School of Physics, Trinity College Dublin, Dublin 2, Ireland }
\affiliation{$^{5}$Department of Physics and Astronomy, Uppsala University, Regementsvägen 1, Uppsala, SE-752 37, Sweden} 
\affiliation{$^{6}$Department of Physics, Brooklyn College, CUNY, 2900 Bedford Ave., Brooklyn, NY 11210, USA }
\affiliation{$^{7}$The Graduate Center, CUNY, 365 Fifth Avenue, New York, New York 10016, USA }
\date{\today}

\begin{abstract}
We model changes of magnetic ordering in Mn-antiperovskite nitrides driven by biaxial lattice strain at zero and at finite temperature. We employ a non-collinear spin-polarised density functional theory to compare the response of the geometrically frustrated exchange interactions to a tetragonal symmetry breaking (the so called piezomagnetic effect) across a range of Mn$_3$AN (A = Rh, Pd, Ag, Co, Ni, Zn, Ga, In, Sn) at zero temperature. Building on the robustness of the effect we focus on Mn$_3$GaN and extend our study to finite temperature using the “disordered local moment” (DLM) first-principles electronic structure theory to model the interplay between the ordering of Mn magnetic moments and itinerant electron states. We discover a rich temperature-strain magnetic phase diagram with two previously unreported phases stabilised by strains larger than 0.75\% and with transition temperatures strongly dependent on strain. We propose an elastocaloric cooling cycle crossing two of the available phase transitions to achieve simultaneously a large isothermal entropy change (due to the first order transition) and a large adiabatic temperature change (due to the second order transition).
\end{abstract}

\maketitle

%******************************************************************************
\section{Introduction}\label{se:intro}
Large magnetocaloric effects (MCE) are available in materials with strong electronic correlations such as Gd$_5$Si$_2$Ge$_2$,\cite{pecharsky1997giant} LaFe$_{13-x}$Si$_x$-based alloys,\cite{shen2009recent} MnFeP$_{0.45}$As$_{0.55}$\cite{tegus2002transition} or Ni-Mn-Sn alloys.\cite{krenke2005inverse} 
The reliance on rare earth based permanent magnets to drive magnetic phase transitions in materials listed above limits the potential of magnetocaloric cooling to replace the current technology based on vapour compression. 
Therefore, using  lattice strains to induce a large entropy change at room temperature promises to open new pathways to energy efficient solid state cooling. 
Large mechanocaloric effects have been demonstrated in shape memory alloys, e.g., elastocaloric effect (eCE) in Ni-Ti\cite{cui2012demonstration,tuvsek2015elastocaloric} or Cu-Zn-Al,\cite{bonnot2008elastocaloric} and barocaloric effect (BCE) in Ni-Mn-In\citep{manosa2010giant}.
A broadening of the usable temperature range by strain has been proposed in Ni-Mn-Ga-Co films.\cite{schleicher2015epitaxial}

Magnetic transitions driven by lattice strains have been reported in several perovskite oxides. Ferromagnetic
(FM) and G-type antiferromagnetic (AFM) phases have been observed in SrCoO$_{3-\delta}$ films subject to low (SrTiO$_3$ substrate) and large (DyScO$_3$ substrate) tensile epitaxial strains, respectively\cite{callori2015strain}.
An increase (decrease) of N\'eel temperature due to compressive (tensile) biaxial strain was predicted in AFM SrTcO$_3$ films\cite{ma2015strain} 
%($T_N$ was inferred assuming the mean field approximation from exchange integrals obtained by fitting of total energies calculated $ab~initio$ zero temperature for different AFM ordering to a Heisenberg model).
A very strong dependence of $T_N$ on biaxial strain ($\approx50$~K per 1\%) has been predicted\cite{lee2010epitaxial} and subsequently confirmed experimentally\cite{maurel2015nature} for G-type AFM phase of SrMnO$_3$. 
The ability to drive a magnetic phase transition with a large entropy change, $\Delta S = 9$~J/kgK, by means of biaxial strain was demonstrated in La$_{0.7}$Ca$_{0.3}$MnO$_3$ film on BaTiO$_3$ substrate.\cite{moya2013giant}
Ferrielectric ammonium sulphate\cite{lloveras2015giant} and spin crossover (SCO) materials\cite{Sandeman2016Research} have also been proposed as new classes of mechanocaloric materials recently. 

Here we study elastocaloric properties of Mn-based antiperovskite nitrides with frustrated non-collinear magnetic structure. This family of materials was first examined in 1970s.\cite{fruchart1971structure,fruchart1978magnetic} The last 10 years have seen renewed interest in these metallic compounds fuelled by a demonstration of large negative thermal expansion (NTE) in Mn$_3$AN (A = Ga, Zn, Cu)\cite{takenaka2005giant} at the first order phase transition to a paramagnetic (PM) state. More recently NTE was studied also in Mn$_3$NiN (stoichiometric\cite{wu2013magnetic} and doped.\cite{deng2015invar}) The related magnetovolume effect\cite{takenaka2014magnetovolume} was measured systematically in a range of Mn$_3$AN. The peak values were observed in Mn$_3$ZnN and Mn$_3$GaN which is consistent with the large BCE measured in Mn$_3$GaN at $T_N$ = 288~K.\cite{matsunami2014giant}
Our study is further motivated by a successful epitaxial growth of Mn$_3$GaN thin film on ferroelectric perovskite substrates.\cite{tashiro2013preparation}

We start by exploring piezomagnetic effects (PME) across a range of Mn$_3$AN (A = Rh, Pd, Ag, Co, Ni, Zn, Ga, In, Sn) using spin density functional theory (SDFT) at zero temperature, building on our earlier study.\cite{zemen2015piezomagnetic}
The PME is characterized by a linear dependence of the induced net magnetic moment, $M_{net}$, on strain,\cite{lukashev2008theory,zemen2015piezomagnetic} which distinguishes it from the quadratic magnetoelastic effect. 
In Mn$_3$AN the PME originates from geometrically frustrated exchange interactions between three Mn atoms in the unit cell which lead to a strong spin-lattice coupling.\cite{lukashev2008theory,gomonaj1992phenomenologic}
This is in contrast with magnetostriction, widely used in spintronic devices, which is driven by the more subtle relativistic spin-orbit coupling. It is worth highlighting in this context that spintronics and solid state cooling have traditionally focused on FM materials. However, AFMs have received much attention in both fields recently fuelled by significant experimental progress including: the demonstration of a giant barocaloric effect in Mn$_3$GaN mentioned above;\cite{matsunami2014giant} the observation of a large room temperature anomalous Hall effect in Mn$_3$Sn  (with triangular AFM structure as in the Mn-antiperovskite family);\cite{nakatsuji2015large} the switching between two stable collinear AFM states in FeRh;\cite{marti2014room} the detection of an AFM state using tunnelling anisotropic magnetoresistance (TAMR) in Pt/MgO/IrMn tunnel junction;\cite{park2011spin} and even all-electric room-temperature switching and detection of staggered AFM moment direction in a CuMnAs-based memory.\cite{wadley2015electrical} 

After exploring PME at zero temperature, we continue by developing a SDFT-based disordered local moment (DLM) theory for the study of finite temperature effects on the magnetic ordering. We show that the spin-lattice coupling also renders the $T_N$ and the magnetic entropy in Mn$_3$GaN highly sensitive to tetragonal lattice distortions. 
We construct the temperature-strain magnetic phase diagram and associated entropy changes in Mn$_3$GaN where a giant BCE at a AFM-PM transition has been observed recently. \cite{matsunami2014giant}
We also discover a collinear AFM and a collinear ferrimagnetic (FIM) phase stabilised by tensile and compressive strain, respectively. 
Both phases are separated by first and second-order transitions from the triangular AFM and PM states. This exceptional phase diagram allows us to design an elastocaloric cooling cycle combining a large isothermal entropy change (due to the abrupt phase transition) and a large adiabatic temperature change (due to the gradual phase transition at a critical temperature strongly dependent on the stimulus\cite{sandeman2012magnetocaloric} - the biaxial strain). This mechanism contrasts with the elastocaloric cycles based on a single phase transition in La$_{0.7}$Ca$_{0.3}$MnO$_3$\cite{moya2013giant} or in shape memory alloys.\cite{tuvsek2015elastocaloric,bonnot2008elastocaloric}
Moreover, the transition temperatures in Mn$_3$GaN are in the room temperature range and can be further tuned by partially substituting atom A by an element with a different number of valence $s$- and $p$-electrons.\cite{fruchart1978magnetic,takenaka2014magnetovolume}

\subsection{Magnetic structure}

\begin{figure}
\includegraphics[width=0.97\columnwidth]{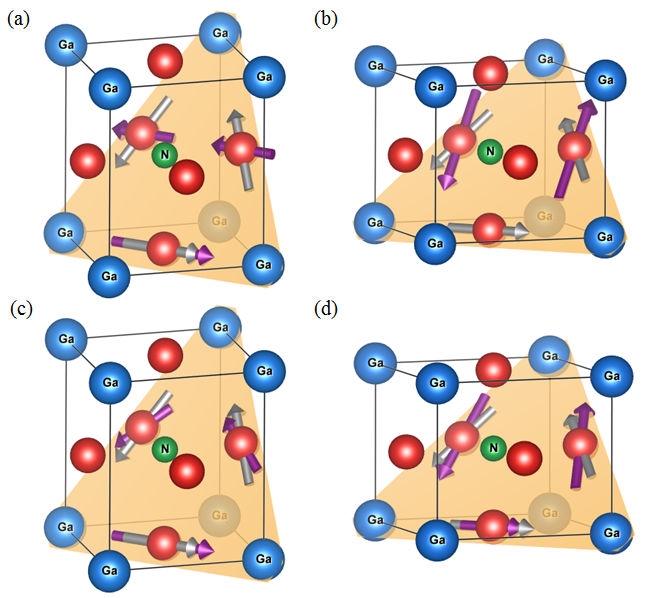}
\caption{(Color online) The strained Mn-antiperovskite structure including the local magnetic moments on Mn sites - silver arrows show the triangular AFM state for a lattice with cubic symmetry; magenta arrows represent the piezomagnetic response of Mn$_3$GaN to compressive (a,c) and tensile (b,d) strain at zero (c,d)\cite{zemen2015piezomagnetic} and finite (a,b) temperature, canting and changes of size are not to scale: (a)~Collinear ferrimagnetic, $M_{net}\parallel [110]$; (b)~Collinear AFM phase, $M_{net}=0$; (c)~Canted triangular phase, $M_{net}\parallel [\bar{1}\bar{1}0]$; (d)~Canted triangular phase, $M_{net}\parallel [110]$ axis;}
\label{f_structure}
\end{figure}

%%%%%%%%%% results T=0 %%%%%%%%%%%%%%%%%%%%%%
Fig.~\ref{f_structure} shows the distorted unit cell of Mn$_3$GaN as an example of the non-collinear magnetic structure of Mn$_3$AN. In the ground state (with cubic lattice) the fully compensated AFM magnetic structure corresponds to the $\Gamma^{5g}$ representation\citep{fruchart1978magnetic,bertaut1968diffraction} indicated by silver arrows. The antiferromagnetic exchange coupling between the neighbouring Mn atoms leads to the frustration (in the triangular lattice of (111) plane, orange online). The three equal-sized local magnetic moments on the Mn sites have an angle of $2\pi/3$ between their directions. 
Another type of AFM ordering ($\Gamma^{4g}$ representation) occurring in Mn$_3$AN (e.g. Mn$_3$SnN) is obtained after a simultaneous rotation of all three local magnetic moments by $\pi/2$ within the (111) plane. The chirality is the same as in case of $\Gamma^{5g}$ but the local moments all point inside (outside) the triangle of (111) plane.\citep{fruchart1971structure} The energy difference between $\Gamma^{4g}$ and $\Gamma^{5g}$ ordering is purely due to the spin-orbit coupling whereas the non-collinearity and magneto-structural coupling is due to the exchange interaction.

An applied biaxial strain $e_{xx} = e_{yy} = (a-a_0)/a_0 \neq e_{zz}$ (where $a_0$ is a lattice parameter of the relaxed structure) relieves the frustration which leads to canting and relative change of size of local moments. 
The "canted triangular" state is represented by magenta arrows in Fig.~\ref{f_structure} (c,d).
% begin from PME
Both effects contribute to an induced net moment,
\begin{equation}
M_{net} \equiv 2 M_1 \cos(2\pi/3+\theta_1)+M_3,
\label{MM}
\end{equation}
which in case of Mn$_3$GaN is (anti)parallel to the $[110]$ axis for (compressive) tensile strain. 
$M_1 = M_2 \ne M_3$ are the magnitudes of the local magnetic moments and $\theta_1 = - \theta_2$ are canted angles with respect to the ground state triangular order. 
The Mn moment at the bottom of the unit cell, $M_3$, is parallel to the $[110]$ crystal axis and does not cant ($\theta_3 = 0$).

\begin{figure}
\includegraphics[width=0.97\columnwidth]{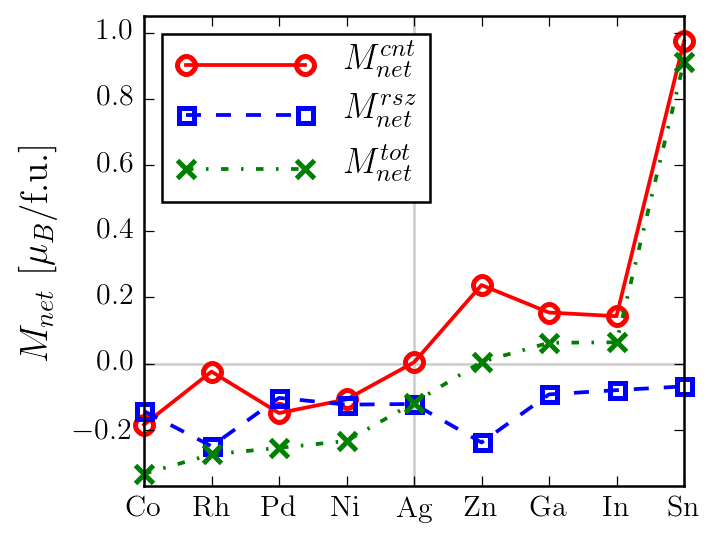}
\caption{(Color online) Comparison of the piezomagnetic effect in nine Mn$_3$AN materials (A labels the x-axis): Contributions to $M^{tot}_{net}$ (denoted as $M_{net}$ in the text) from Mn local moment canting and resizing induced by tensile stain, $\varepsilon_{xx}=1\%$ are shown. (We have reported the total moment, $M_{net}$, earlier.\cite{zemen2015piezomagnetic})}
\label{f_pme}
\end{figure}

\section{Piezomagnetic Effect}\label{se:pme}

We model the piezomagnetic effect across a range of Mn$_3$AN materials at zero temperature using SDFT. We employ the projector augmented-wave (PAW) method implemented in VASP code\cite{kresse1999ultrasoft} within the Perdew-Burke-Ernzerhof (PBE) generalized gradient approximation.\citep{perdew1996generalized}
This code allows for relaxation of fully unconstrained noncollinear magnetic structures.\citep{hobbs2000fully} We use a 12x12x12 k-point sampling in the self-consistent cycle. The cutoff energy is 400~eV. The Mn local magnetic moments are evaluated in atomic spheres with the default Wigner Seitz radius.\cite{zemen2015piezomagnetic}

Fig.~\ref{f_pme} shows the two contributions to the PME separately for the nine Mn$_3$AN's with tensile strain $\varepsilon_{xx} = 1\%$: 
\begin{eqnarray}
M^{rsz}_{net} & \equiv & 2 M_1 \cos(2\pi/3)+M_3 = M_3-M_1, \\
M^{cnt}_{net} & \equiv & 2 M_0 \cos(2\pi/3+\theta_1)+M_0 \label{eq:cnt},\;
\end{eqnarray}
where $M^{rsz}_{net}$ is the net moment due to the change of size of the local magnetic moments, $M^{cnt}_{net}$ is the net moment due to their canting, and $M_0$ is the moment common to all Mn atoms at zero strain. $M^{rsz}_{net}$ is negative for all systems which reveals the universal effect of increasing Mn moment size with increasing distance to the nearest nitrogen. (Results in Fig.~\ref{f_pme} assume unit cell volume conservation when Mn$_3$ is closer to nitrogen than Mn$_1$ for tensile strain.) On the other hand, $M^{cnt}_{net}$ is negative (positive canting, $\theta_1 > 0$) when atom A is a transition metal, except Ag and Zn, and positive (negative canting, $\theta_1 < 0$) for the rest. Both Ag and Zn have a fully filled $3d$-band well below the Fermi energy ($E_F$) so their hybridization with Mn $3d$-states forming a wider band around $E_F$ is similar to Ga or In which explains its positive $M^{cnt}_{net}$. Therefore we conclude that when atom A has only $s$ or $p$-states relatively close to $E_F$ (within 4~eV) then the canted angle is negative, $\theta_1 < 0$. In case of smaller energy separations of $d$-states of atom A from Mn $d$-states the canted angle is positive, $\theta_1 > 0$.

It is remarkable that Mn$_3$ZnN has one of the largest $M^{cnt}_{net}$ due to moment resizing which gets almost completely compensated by the canting. In other words, its magnetic system is very sensitive to the tetragonal distortion but a measurement of $M_{net}$ would not reveal that. The canting in case of Ag is very weak and hard to resolve numerically. At the same time, 
Mn$_3$SnN combines a very large PME with $T_N = 475$~K\cite{LB1981} so it has a potential for spintronic applications  including magnetic memory, magnetic sensors, and actuators.\cite{zemen2015piezomagnetic}

A table summarizing the equilibrium lattice parameters, the size of the Mn local moments, or the Poisson's ratios obtained by SDFT calculations (VASP) can be found in our previous work on PME.\cite{zemen2015piezomagnetic} It should be noted that we explore the response of the magnetic order to strain only in small surroundings of the unstrained ground state. We use PME as a measure of the frustration of the exchange interaction across the material series. We exploit the fact that the triangular AFM structure is predicted to be a local energy minimum in our SDFT model even though it has not been confirmed experimentally in Mn$_3$AN (A = Ag, Co, Pr, Rh).

\section{\textit{Ab-initio} Theory of geometrically frustrated magnetic transitions and caloric responses} 
\label{se:theory}

Having established the robustness of the piezomagnetic effect across a range of Mn$_3$AN systems and its dependence on the valence states of atom A, we proceed to investigate the non-collinear magnetism and the caloric properties of these systems at finite temperatures. We focus on strain-induced caloric responses which arise from distortions of the magnetic structure. To achieve this we have extended the SDFT-based DLM theory which describes self-consistently the interplay between disordered local moments (DLMs) and itinerant electrons in magnetic metals at finite temperature.\cite{gyorffy1985first}

\subsection{Disordered local moment modelling and the free energy}
Our modelling of a magnetic system subject to a strain $\bm{\varepsilon} \equiv \varepsilon_{\alpha\beta}$ is analogous to earlier studies of magnetic phase transitions and associated magnetic field-induced entropy changes in FeRh,\cite{staunton2014fluctuating} some relevant Gadolinium intermetalics,\cite{petit2015complex} and the heavy rare earth elements.\cite{mendive2017HRE} The model assumes a time-scale separation between the slowly varying orientations of the local moments of Mn atoms, and the remaining faster electronic degrees of freedom. We label the local moment orientations by local spin polarization axes $\{\hat{\bm{e}}_i\}$ fixed to each Mn atom. Under these circumstances we can evaluate a generalized electronic grand potential $\Omega(\lbrace \hat{\bm{e}}_i \rbrace,\bm{\varepsilon})$ using SDFT,\cite{gyorffy1985first} with the spin density constrained to the local moment configuration $\lbrace \hat{\bm{e}}_i \rbrace$. The probability of each configuration $\{\hat{\bm{e}}_i\}$ is then calculated as:
\begin{equation}
P(\lbrace \hat{\bm{e}}_i \rbrace) = \exp[-\beta\Omega(\lbrace \hat{\bm{e}}_i \rbrace,\bm{\varepsilon})]/Z=
\prod_i{P_i(\hat{\bm{e}}_i)},
\label{PROB}
\end{equation}
where $Z$ is the constrained partition function, $1/\beta = k_\text{B}T$ ($k_\text{B}$ being the Boltzmann constant), and the single site probabilities $P_i(\hat{\bm{e}}_i)$ are calculated within a mean field approach in terms of the Weiss fields $\{\bm{h}_i\}$,
\begin{equation}
P_i(\hat{\bm{e}}_i)=\frac{\exp\left[\beta\bm{h}_i\cdot\hat{\bm{e}}_i\right]}{\int{\text{d} \hat{e}_i\exp\left[\beta\bm{h}_i\cdot\hat{\bm{e}}_i\right]}}.
\label{PROB1}
\end{equation}

An ensemble average over non-collinear local moment configurations is carried out within the coherent potential approximation (CPA)\cite{stocks1978complete,johnson1986density,ebert2011calculating} framework to find a specific magnetic state of the system. This state is specified by magnetic order parameters:
\begin{equation}
\left\{ \bm{m}_i = \int \hat{\bm{e}}_i P_i( \hat{\bm{e}}_i ) \text{d}\hat{\bm{e}}_i = \left[\frac{-1}{\beta h_i}+\coth(\beta h_i)\right]\hat{h}_i \right\},
\label{OP}
\end{equation}
where $h_i = |\bm{h}_i|$ and $\bm{h}_i = h_i\hat{h}_i$. The magnitudes $m_i = |\bm{m}_i|$ describe the magnetic disorder of the DLMs associated to each magnetic moment at each Mn atom. They range from 0 for the high-temperature fully disordered PM state to 1 for the fully ordered magnetic structures (including triangular AFM) at $T = 0$~K. The local moments $M_i = \mu_i m_i$ on Mn sites are related to the order parameters via a set of local moment sizes $\mu_i$ determined by the generalised SDFT.\cite{gyorffy1985first}

The free energy can be written as a function of these magnetic order parameters, lattice strain, and temperature
\begin{equation}
\mathcal{F}(\lbrace \bm{m}_i \rbrace,\bm{\varepsilon},T) = \bar{\Omega}(\lbrace \bm{m}_i \rbrace,\bm{\varepsilon}) - T\bar{S}_{mag},
\label{Efree}
\end{equation}
where $\bar{\Omega}= \langle \Omega(\lbrace \hat{\bm{e}}_i \rbrace,\bm{\varepsilon}) \rangle_{\lbrace \bm{m}_i \rbrace}$ is the SDFT-based internal energy averaged over orientations of the DLMs and consistent with the constrained system described by $\lbrace \bm{m}_i \rbrace$, and $\bar{S}_{mag}$ is the magnetic entropy contribution to the total entropy $S_{tot}$. $\bar{S}_{mag}$ can be easily calculated by performing the integral
\begin{equation} \label{Smag}
\bar{S}_{mag} = - k_\text{B}T \sum_i \int P_i(\hat{\bm{e}}_i) \mathrm{ln} P_i(\hat{\bm{e}}_i)\text{d}\hat{\bm{e}}_i,
\end{equation}
such that it only depends on the quantities $\{\beta\bm{h}_i\}$ (or $\{\bm{m}_i\}$).
The electronic entropy contribution $\bar{S}_{el}$ is contained in $\bar{\Omega}$.\cite{Mermin1965,staunton2014fluctuating} We calculate $\bar{S}_{el}$ from the Sommerfeld expansion, $\bar{S}_{el} \approx \frac{\pi^2}{3} k_\text{B}^2 T\bar{n}(E_F)$, where $\bar{n}(E_F)$ is the electronic density of states at Fermi energy available from the SDFT and averaged over all local moment orientations.\cite{staunton2014fluctuating}
%In our calculations we have found that $\bar{S}_{el}$ is negligible compared to $\bar{S}_{mag}$ in Mn$_3$GaN.

\subsection{Magnetic phase diagram}

Minimization of the free energy of eq.~(\ref{Efree}) with respect to the order parameters $\lbrace \bm{m}_i \rbrace$ in the absence of external magnetic field leads to an expression for the Weiss field at the atomic site $i$\cite{gyorffy1985first}
\begin{equation}
\bm{h}_i = - \nabla_{\bm{m}_i}\bar{\Omega}(\{\bm{m}_i\},\bm{\varepsilon}).
\label{WF}
\end{equation}
We can see from Eq.~(\ref{OP}) that the Weiss fields divided by temperature, $\{\beta\bm{h}_i\}$, in turn determine the order parameters $\{\bm{m}_i\}$. This provides a basis for a self-consistent calculation of a stable magnetic order $\lbrace \bm{m}_i \rbrace$ for a given temperature and lattice parameters (strain).\cite{staunton2014fluctuating,mendive2017HRE}
In general, several solutions, $\lbrace \bm{m}_i^{(1)} \rbrace$, $\lbrace \bm{m}_i^{(2)} \rbrace$, $\dots$ may be found at different local minima of the free energy.
These competing local minima can be tracked across a range of temperatures and strains and a transition temperature is defined by a switching of the global minimum from, e.g., $\mathcal{F}(\lbrace \bm{m}_i^{(1)}\rbrace,\bm{\varepsilon},T)$ to a new global minimum $\mathcal{F}(\lbrace \bm{m}_i^{(2)}\rbrace,\bm{\varepsilon},T)$.

%Firstly, an initial guess of $\lbrace \bm{m}_i \rbrace$ is chosen and then the Weiss fields $\{\bm{h}_i\}$ are calculated from Eq.\ (\ref{WF}). Then $\lbrace \bm{m}_i \rbrace$ are calculated again and compared with the initial guess. This process is repeated iteratively until convergence is achieved.

%We remark that the resulting magnetic phase diagram was not very dependent on the initial guess of $\{\bm{m}_i\}$. 

%The self consistent calculation of a magnetic structure minimizing the free energy described above requires $\{\bm{h}_i\}$ to be calculated. 
In order to find stable magnetic states on a dense temperature-strain grid and to compare the corresponding free energies we write our internal energy $\bar{\Omega}(\lbrace \bm{m}_i \rbrace,\bm{\varepsilon})$ as an expansion in powers of temperature-dependent parameters $\{\bm{m}_i\}$ with strain-dependent coefficients (examples are given in Sec.~\ref{se:eCE}). We obtain an analytical expression for $\nabla_{\bm{m}_i}\bar{\Omega}$ of Eq.~(\ref{WF}) which can be calculated explicitly within our SDFT-DLM theory.\cite{gyorffy1985first} Then the required temperature-independent expansion coefficients can be calculated \textit{ab-initio} in the following way: (i) We start by the direct calculation of $\nabla_{\bm{m}_i}\bar{\Omega}$ for a sufficiently large set of magnetic configurations $\lbrace \bm{m}_i \rbrace$; (ii) The coefficients are extracted by fitting these values to the analytical expression for $\nabla_{\bm{m}_i}\bar{\Omega}$ for a given value of strain. (iii) This is repeated for different lattice parameters in order to determine the dependence of the expansion coefficients on strain (e.g., in case of Mn$_3$GaN we find that the quadratic coefficients can be fitted to a linear strain-dependence.) Once the coefficients including the strain-dependence are obtained the self-consistent calculation of the stable magnetic states at each point of the temperature-strain phase diagram can ensue without the computationally expensive evaluation of $\nabla_{\bm{m}_i}\bar{\Omega}$ from SDFT-DLM. Moreover, the free energy can be evaluated from the same expansion coefficients in order to identify the global energy minima at each point of the phase diagram. 

It should be noted that the choice of configurations $\lbrace \bm{m}_i \rbrace$ used to initialize the self-consistent cycle of Eqs.~(\ref{OP}) and~(\ref{WF}) is guided by instabilities of the high temperature PM state identified by the application of DLM linear response theory\cite{gyorffy1985first,Hughes2007} (see appendix \ref{DLMlinresp} for details). In the case of Mn$_3$GaN we found that the triangular and collinear magnetic perturbations lead to the strongest response just below the N\'eel temperature at zero and at high enough biaxial strain, respectively. The stability of the corresponding magnetic phases (shown in Fig.~\ref{f_structure}) well below the transition temperature was confirmed by the rapidly converging self-consistent calculation ($\approx 20$ iterations).

\subsection{Calculation of caloric responses}

The cooling capacity and the temperature span are the key characteristics of refrigerants and pertinent cooling cycles. The adiabatic temperature change ($\Delta T_{ad}$) and the isothermal entropy change ($\Delta S_{iso}$) induced by the application and/or removal of an external field are directly related to these characteristics and are typically used to compare refrigerants.\cite{sandeman2012magnetocaloric,gschneidner2000magnetocaloric,moya2014caloric}
In principle the lattice vibrations could be incorporated self-consistently within our SDFT-DLM~\cite{ebert2015calculating} theory and a direct magneto-phonon coupling be obtained. However the entropy of lattice vibrations cannot be obtained within the SDFT-DLM at present.\cite{ebert2015calculating} As we are interested in the calculation of the adiabatic temperature change $\Delta T_{ad}$ the incorporation of the lattice vibrations acting as a thermal bath (or reservoir) is fundamental to avoid unphysical results.\cite{mendive2015magnetocaloric} We have consequently implemented a standard simple Debye model for the vibrational entropy\cite{Oliveira2010}
\begin{equation}
S_{vib} =  k_\text{B} \Biggl[-3\ln\left(1-e^{-\frac{T}{\theta_\text{D}}}\right)+12\left(\frac{T}{\theta_\text{D}}\right)^{3}\int_{0}^{\frac{\theta_\text{D}}{T}}\frac{x^{3}}{e^{x}-1}dx\Biggr],
\label{EQTS}
\end{equation}
where $\theta_\text{D}$ is the Debye temperature (see Appendix A for further details). Note that the resulting vibrational entropy does not depend on strain or volume of the unit cell.
%we assume a Debye temperature of 429.2~K, which becomes after rescaling to our transition temperature $\theta_D = 452.2$~K.

Here we are interested in evaluating $\Delta S_{iso}$ and $\Delta T_{ad}$ when they are induced by biaxial strain application. For a finite change of the strain ($\bm{\varepsilon}_0\rightarrow\bm{\varepsilon}_1$) we can calculate $\Delta S_{iso}(T,\bm{\varepsilon}_0\rightarrow\bm{\varepsilon}_1)$ at temperature $T$ as
\begin{equation}
\Delta S_{iso}(T,\bm{\varepsilon}_0\rightarrow\bm{\varepsilon}_1)=S_{tot}(\bm{\varepsilon}_1,T)-S_{tot}(\bm{\varepsilon}_0,T),
\label{Siso1}
\end{equation}
while $\Delta T_{ad}(T,\bm{\varepsilon}_0\rightarrow\bm{\varepsilon}_1)$ can be estimated from
\begin{equation}
S_{tot}(T,\bm{\varepsilon}_0)=S_{tot}(T+\Delta T_{ad},\bm{\varepsilon}_1),
\label{Siso2}
\end{equation}
where the total entropy is $S_{tot}=\bar{S}_{mag}+S_{vib}+\bar{S}_{el}$. We note that the entropy changes originate from the change of the geometrically frustrated magnetic ordering induced by the application of mechanical stress on the lattice system and therefore a strong spin-lattice coupling is necessary.

\section{The Elastocaloric Effect} \label{se:eCE}

In this section we implement our SDFT-based DLM theory to explore the geometrically frustrated  non-collinear magnetism in strained Mn$_3$GaN at finite temperature. This is motivated by the recent observation of a large barocaloric effect\cite{matsunami2014giant} as well as our SDFT simulations at zero temperature in Sec.~\ref{se:pme}. Our DLM method is well suited for metallic Mn$_3$GaN where the local magnetic moments are relatively well localized\cite{lukashev2010spin} while their interaction with the delocalized gallium $p$-states determines the size\cite{zemen2015piezomagnetic} and direction of the strain-induced $M_{net}$ as shown in Fig.~\ref{f_pme}. We start by comparing our results with available experimental data for the unstrained cubic system. Then we apply our theory to biaxial strain and study the elastocaloric effect. 

Note that in the following we provide analytical expressions for $\bar{\Omega}(\{\bm{m}\}_i,\bm{\varepsilon})$ that capture satisfactorily our SDFT-DLM results for $\{\bm{h}_i = - \nabla_{\bm{m}_i}\bar{\Omega}\}$ and extract the coefficients involved. The minimization of the free energy, the calculation of the stable magnetic configurations and caloric responses, and the construction of the magnetic phase diagrams are performed as described in Sec.~\ref{se:theory}.

\subsection{Unstrained cubic system}\label{se:isotropic}

Our DLM model predicts the triangular AFM order of Fig.~\ref{f_structure} as the most stable structure in agreement with early neutron diffraction studies\cite{bertaut1968diffraction,fruchart1978magnetic,fruchart1971structure} and with our zero-temperature simulations of sec.~\ref{se:pme}.
In the case of Mn$_3$GaN with an unstrained lattice (cubic symmetry) all 3 Mn atoms can be described by a single order parameter $m$ even though the angle between the spin polarization axes $\hat{\bm{e}}_i$ is $2\pi/3$ between each pair. Thus $m$  is the common length of the three order parameters. We find that our SDFT-DLM internal energy can be approximated satisfactorily by a two term expansion:
\begin{equation} \label{FreeEcub}
\bar{\Omega}(\{\bm{m}_i\},\bm{\varepsilon}) = am^2 + bm^4,
\end{equation}
where the coefficients $a = -31.47$~meV and $b = -33.11$~meV have been obtained by fitting for relaxed Mn$_3$GaN.

At the N\'eel temperature the negative quartic coefficient $b$ is larger than its positive counterpart in the expansion of magnetic entropy in powers of the order parameter (see eq.~(\ref{Efree})) which indicates a first-order phase transition between the triangular AFM and PM state in agreement with experiment.\cite{matsunami2014giant} We find a transition temperature $T_N=304$~K which is very close to reported experimental values $T_N=288$~K\cite{takenaka2014magnetovolume} and 290~K.\cite{matsunami2014giant}
When we repeat our simulation for a slightly larger lattice parameter (preserving the cubic symmetry and the form of eq.~(\ref{Efree})), we find an increase of $T_N$ with increasing unit cell volume in agreement with experiment.\cite{matsunami2014giant} Matsunami \textit{et al.} have recently measured a very large isothermal entropy change $\Delta S = 22.3$~J/kgK at the AFM-PM transition in Mn$_3$GaN\cite{matsunami2014giant} which correlates with the large magnetovolume effect\cite{takenaka2014magnetovolume} driven by the geometric frustration and the abrupt change of effective amplitudes of Mn magnetic moments.
Our DLM theory for unstrained cubic Mn$_3$GaN finds a significantly larger BCE (Fig.~2 of Ref.~\citep{matsunami2014giant}) as the calculated magnetic entropy change at $T_N$ is $\bar{S}_{mag}(T_N+\delta T)-\bar{S}_{mag}(T_N-\delta T) = \Delta\bar{S}_{mag} = 103.2$~J/kgK, accompanied by a large change of the magnetic order parameter $\Delta m = 0.74$. The electronic entropy change is very low, $\Delta \bar{S}_{el} = 0.035$~J/kgK, and our $\Delta S_{vib}$ (we have assumed a Debye temperature of 429.2~K, which becomes $\theta_D = 452.2$~K after rescaling to our transition temperature\cite{takenaka2014magnetovolume}) %neglecting the magneto-phonon coupling 
also cannot compensate the discrepancy between $\Delta \bar{S}_{mag}$ and the measured $\Delta S_{tot}$.\cite{matsunami2014giant} However, it should be noted that our $\Delta \bar{S}_{mag}$ falls well below the theoretical upper limit  proportional to $k_\text{B}\ln(2J+1)$ which is $161.52$~J/kgK for Mn$_3$GaN, where $J$ is the total angular momentum of the magnetic atom.
%\cite{pecharsky1997giant} 
%S=6.02214179e23*1.3806488e-23/248.5438*1000*ln(2*2+1)*3
At the same time, strong dependence of magnetic transitions on compositional disorder has been shown in Mn$_3$AN\cite{takenaka2014magnetovolume,takenaka2005giant}
and FeRh.\cite{staunton2014fluctuating}
Owing to its ties to the geometric frustration, $\Delta \bar{S}_{mag}$ is likely to be sensitive also to any symmetry lowering due to structural defects in sintered polycrystalline samples. 
Therefore, $\Delta \bar{S}_{mag}$ calculated in a system with perfect stoichiometry and lattice symmetry, hence with a very sharp phase transition, should be regarded as an upper estimate of the entropy change measured at a smoother phase transition in a real sample. This is consistent with our overestimate of $\Delta \bar{S}_{mag}$ and indicates that the simulated material has a sharper phase transition than the available sample.

\subsection{Biaxial strain}\label{se:biaxial}
%In order to describe the strained Mn$_3$GaN at finite temperature (tetragonal lattice) we define three independent order parameters, $\lbrace \bm{m}_1,\bm{m}_2, \bm{m}_3 \rbrace$. As a result, a more complicated expansion of the free energy is found to fit the SDFT result. However, only quadratic and quartic terms involved as in eq.~(\ref{FreeEcub}) across the explored range of strain, $\varepsilon_{xx} \in \langle -2.5, 3 \rangle \%$ within the relative error of the fitting, $1\%$. 
%Their dependence on strain is again linear: strong in case of the quadratic and weak in case of the quartic coefficients. Please see the supplementary information for the full expression and a table of fitted values.

\begin{figure}
\includegraphics[width=1.1\columnwidth]{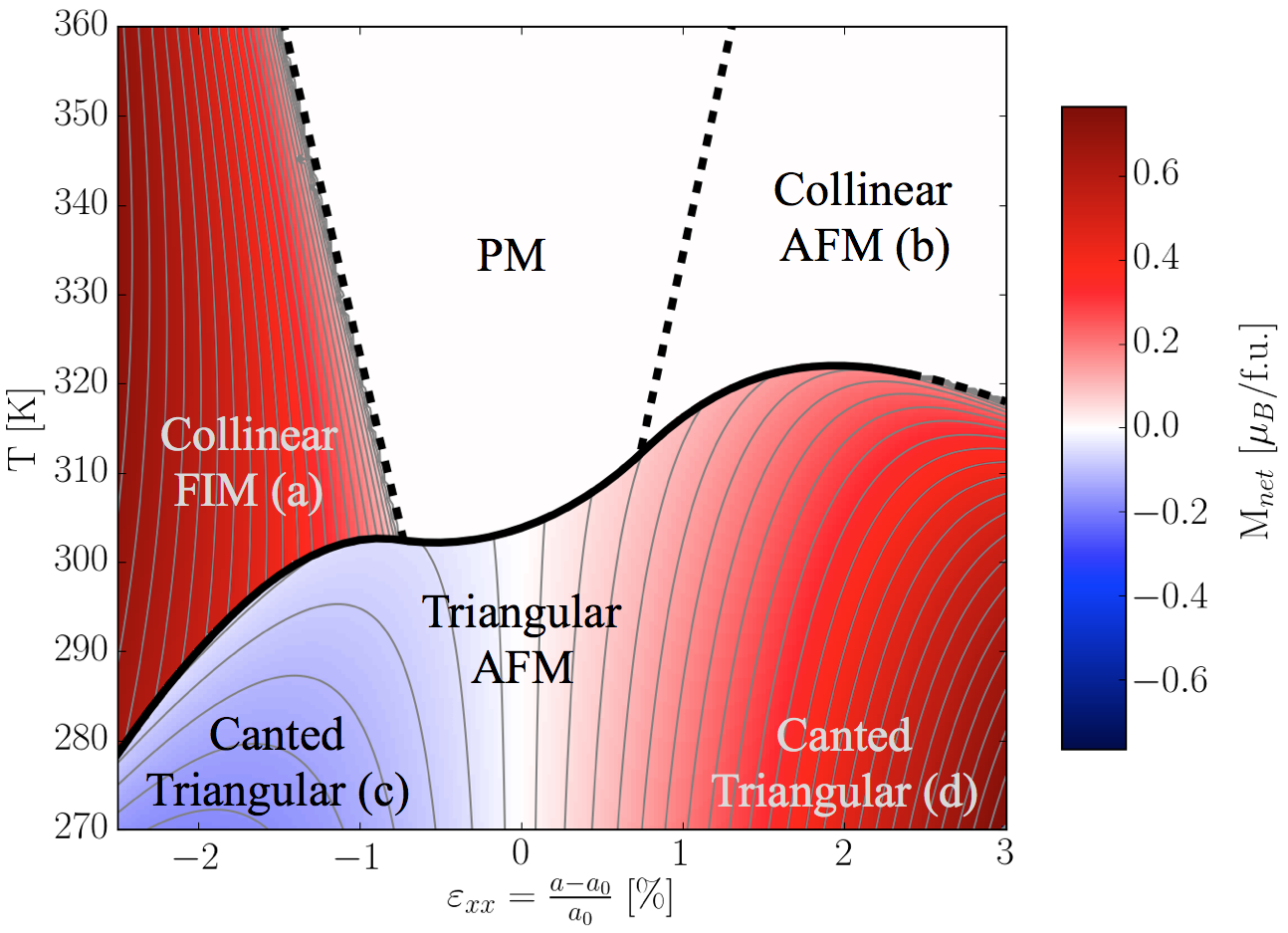}
\caption{(Color online) Magnetic phase diagram for Mn$_3$GaN; colors encodes the size and orientation of the induced moment, $M_{net}>0$ is along the $[110]$ axis; thick black lines mark the first-order (solid) and second-order (dashed) magnetic phase transitions; letters in brackets link to panels of Fig.~\ref{f_structure}.}
\label{f_PD}
\end{figure}

Having compared the results of our DLM modelling to available experimental data for Mn$_3$GaN, we now focus on the effect of biaxial strain. We assume only volume-conserving strains, i.e., Poisson's ratio = 0.5. 
Due to the lower symmetry we have to define three independent order parameters $\lbrace \bm{m}_1,\bm{m}_2, \bm{m}_3 \rbrace$ corresponding to each magnetic moment in Mn within the unit cell. This results in a more complicated expansion of the SDFT- internal energy:
\begin{eqnarray}
\label{FreeEtet}
& & \bar{\Omega}(\{\bm{m}_i\},\bm{\varepsilon}) = \nonumber \\
& & -a_1(m_1^2+m_2^2)-a_2m_3^2-\alpha_1\textbf{m}_3\cdot(\textbf{m}_1+\textbf{m}_2) \nonumber \\
                         & & -\alpha_2\textbf{m}_1\cdot\textbf{m}_2-b_1(m_1^4+m_2^4)-b_2m_3^4 \nonumber \\
                         & & -\beta_1[(\textbf{m}_3\cdot\textbf{m}_1)m_2^2+(\textbf{m}_3\cdot\textbf{m}_2)m_1^2]  \\
                         & & -\beta_2[(\textbf{m}_3\cdot\textbf{m}_1)+(\textbf{m}_3\cdot\textbf{m}_2)](\textbf{m}_1\cdot\textbf{m}_2) \nonumber \\
                         & & -\beta_3(\textbf{m}_1\cdot\textbf{m}_2)m_3^2 \nonumber-\beta_4(\textbf{m}_3\cdot\textbf{m}_1)(\textbf{m}_3\cdot\textbf{m}_2). \nonumber \;
\end{eqnarray}
Including only quadratic and quartic terms in Eq.~(\ref{FreeEtet}) is enough to fit satisfactorily our SDFT-DLM data and capture the relevant physics across the explored range of strain. In order to construct the phase diagrams shown in Figs.~\ref{f_PD}, \ref{f_eCE} and \ref{f_eCE_2D} we have fitted the coefficients of Eq.\ (\ref{FreeEtet}) for 7 different values of $\varepsilon_{xx} \in \langle -1, 1 \rangle \%$ and found a nearly linear dependence of all 10 constants on strain. 
To cover the relevant range of strain $\varepsilon_{xx} \in \langle -2.5, 3 \rangle \%$ we then performed a linear fit across to values of each constant for the 7 available strains. Note that we also extract a linear dependence of the Mn magnetic moments on strain: $M_1 = M_2 = (3.102-0.0361\varepsilon_{xx})\mu_\text{B}$, $M_3 = (3.102+0.0438\varepsilon_{xx})\mu_\text{B}$ for compressive strain and $M_1 = M_2 = (3.102-0.0265\varepsilon_{xx})\mu_\text{B}$, $M_3 = (3.102+0.0444\varepsilon_{xx})\mu_\text{B}$ for tensile strain.
The phase diagram of Fig.~\ref{f_PD} has been constructed by tracking the free energy of competing magnetic phases across the range of strain and temperature with a sufficiently small step allowed by the fitting described above.

We obtained a strong dependence of the quadratic coefficients on the biaxial strain. However, similar changes of the quartic coefficients have negligible impact on the magnetic phase diagram (see appendix B). We therefore concluded that all features of the temperature-strain magnetic phase diagram are determined mainly to two factors: (i) the presence of large quartic coefficients resulting in the first-order nature of the PM-AFM transition at zero strain and (ii) a strong dependence of the quadratic coefficients on $\varepsilon_{xx}$.

%\begin{figure}[h]
%$\includegraphics[width=0.85\columnwidth]{Angles.pdf}
%\caption{(Color online) The canted angle, $\theta_1$, as a function of biaxial strain $\varepsilon_{xx}$ %calculated using our DLM-DFT modelling at $T$=5K.}
%\label{f_PME_angle}
%\end{figure}

At higher temperatures, we predict two novel strain-induced magnetic phases: a collinear (ferrimagnetic) FIM phase for compressive strain $\varepsilon_{xx}<-0.75\%$ shown in Fig.~\ref{f_structure}(a) and a collinear AFM phase for tensile strain $\varepsilon_{xx}>0.75\%$ shown in Fig.~\ref{f_structure}(b). 
Notably, we also determine the order of the phase transitions. Solid (dashed) black lines in Fig.~\ref{f_PD} indicate first (second)-order transitions.
The collinear FIM and AFM emerge from the large change of the quadratic coefficients with $\varepsilon_{xx}$. This is the most conspicuous feature of Fig.~\ref{f_PD} leading to a strong dependence of the second-order transition temperature on strain between these collinear magnetic structures and the PM state.
The transition between canted triangular and collinear AFM states changes from first- to second- order for large tensile strain. There is a tricritical point as a consequence. The color-coding shows $M_{net}$: The collinear AFM state does not possess any net magnetization, whereas the tensile-strained canted triangular and collinear FIM states develop $M_{net}\parallel [110]$ (positive, red) and $M_{net}$ antiparallel to $[110]$ in the compressive-strained canted triangular state (negative, blue). The corresponding induced magnetic field reaches 200~Oe at 1\% strain at room temperature so the strained material has a potential for multicaloric effects (MCE \& eCE). %0.15*0.74*9.274e-24/(3.89e-10^3)*4*pi/1000 (Mn are 0.74 ordered at the transition)

\subsection{Cooling cycles}\label{se:cycle}

\begin{figure}
\includegraphics[width=0.97\columnwidth]{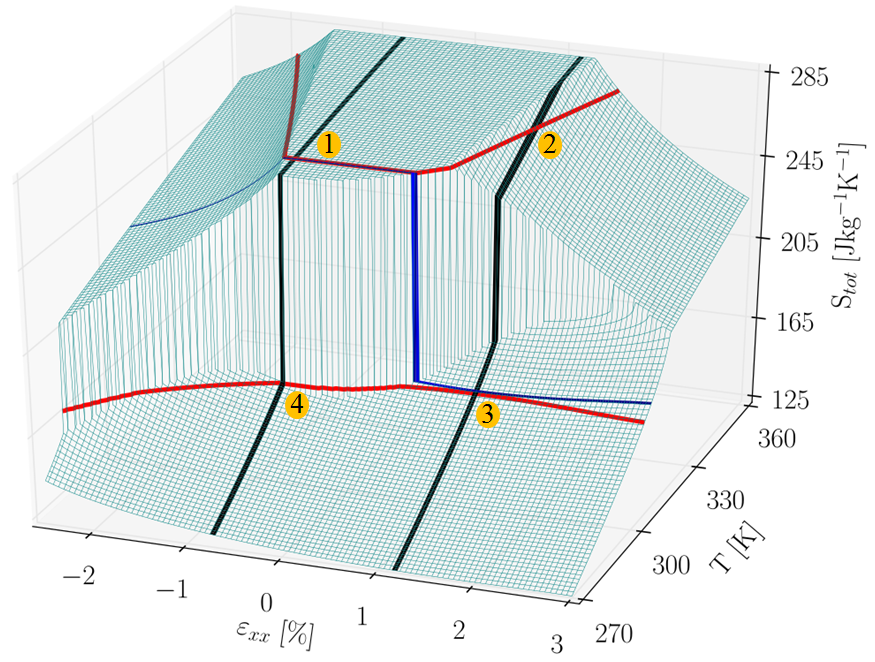}
\caption{(Color online) Total entropy of Mn$_3$GaN; Red contour lines mark adiabatic application of strain at $S_{tot}=170$~and 270~J/kgK; Black lines mark iso-strain cooling ($\varepsilon_{xx}=1.18\%$) and heating ($\varepsilon_{xx}=-0.73\%$); Blue isotherm marks the reference temperature of 308~K; Orange numbers mark the proposed cooling cycle.}
\label{f_eCE}
\end{figure}

Fig.~\ref{f_eCE} presents our magnetic phase diagram from the perspective of total entropy. We note that we have ignored the electronic contribution $S_{el}$ because we have found it to be negligible compared to $\bar{S}_{mag}$ and $S_{vib}$. Fig.~\ref{f_eCE} shows the abrupt entropy change $\Delta S_{tot} \approx 100$~J/kgK at the first-order transition to the PM state at zero strain. The transition gradually becomes less first-order-like as the strain increases and becomes smooth around $\varepsilon_{xx}=2.5\%$ due to the presence of the tricritical point.  
Notably, both the collinear FIM and AFM states at higher temperature show a very strong dependence of $\bar{S}_{mag}$ on strain.

We now propose an elastocaloric cooling cycle utilising the complex pattern of magnetic phase transitions of Fig.~\ref{f_eCE} instead of structural phase transitions of shape memory alloys. The cycle starts by adiabatic application of strain: red line from point (1) to (2), $\bar{S}_{mag}$ decreases at the second-order transition from PM to collinear AFM state which is compensated by an increase in $S_{vib}$ accompanied by a warming of $\approx 25$~K. In the second step, the system is then cooled to its original temperature at constant strain: black line from point (2) to (3), $\bar{S}_{mag}$ further decreases through the first-order transition to the canted triangular state and heat is expelled to the environment. In the third step, a strain is applied adiabatically again: red line from point (3) to (4), $\bar{S}_{mag}$ increases continuously and temperature decreases by $\approx 5$~K. Finally, the refrigerant is warmed up at constant strain: black line from point (4) back to (1), $\bar{S}_{mag}$ increases sharply at the first-order transition from the canted triangular state to PM state and heat is absorbed from the load. 

\begin{figure}
\includegraphics[width=0.97\columnwidth]{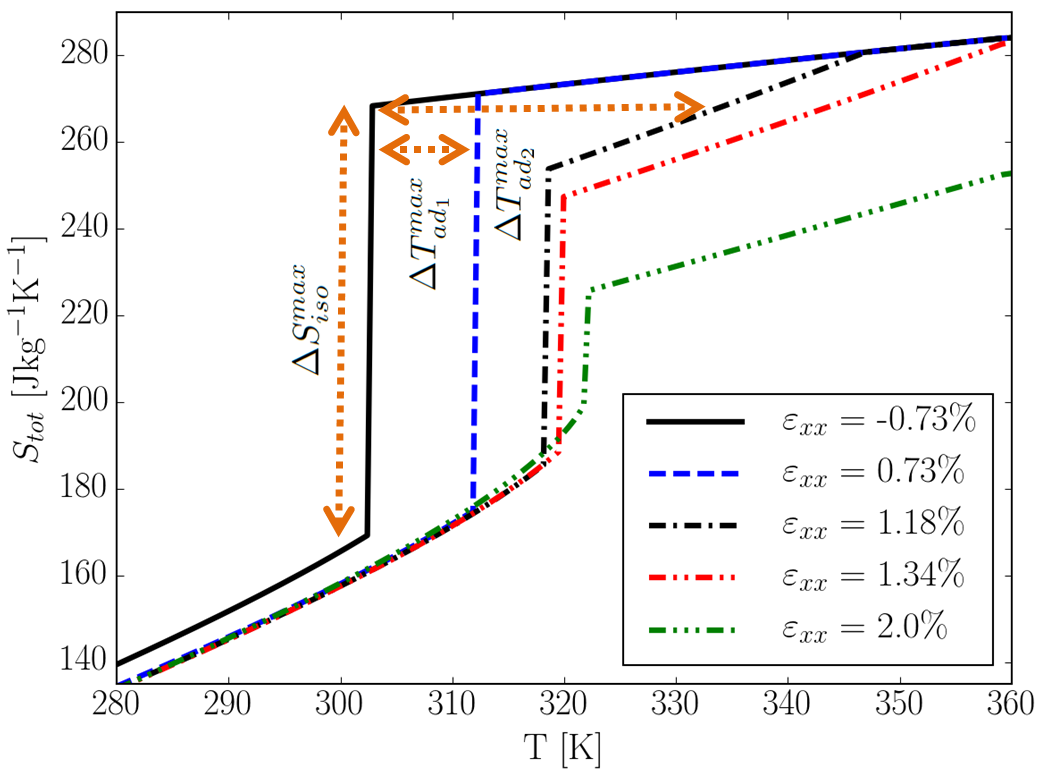}
\caption{(Color online) The total entropy for selected values of strain in Mn$_3$GaN; Black lines correspond to black iso-strain lines in Fig.~\ref{f_eCE}; Blue dashed line crosses only the first-order phase transition (small $\Delta T_{ad_1}^{max}$); All dash-dotted lines cross both the first and second-order transitions and allow for larger $\Delta T_{ad_2}^{max}$.}
\label{f_ST}
\end{figure}

Fig.~\ref{f_ST} shows the dependence of $S_{tot}$ on temperature for five strains which determine $\Delta S^{max}_{iso}$ and different values of $\Delta T_{ad}^{max}$. 
%We estimate $\Delta T_{ad}(\bm{\varepsilon},T)$ from $\bar{S}_{el}(\bm{\varepsilon}_1,T) + 
%\bar{S}_{mag}(\bm{\varepsilon}_1,T) + 
%S_{ph}(\bm{\varepsilon}_1,T) = 
%\bar{S}_{el}(\bm{\varepsilon}_0,T + \Delta T_{ad}) + 
%\bar{S}_{mag}(\bm{\varepsilon}_0,T+\Delta T_{ad}) + 
%S_{ph}(\bm{\varepsilon}_0,T+\Delta T_{ad})$.
We recall that in case of a cooling cycle with a single first-order phase transition driven by external magnetic field, $H_{max}$, using a material with a weak dependence of the Curie temperature on field, $\partial T_C/ \partial H < \sqrt{T/C_p M_{sat}H_{max}}$,\cite{sandeman2012magnetocaloric,zverev2010maximum} its $\Delta T_{ad}^{max}$ cannot reach the highest value allowed by the entropy change, $\Delta T_{ad}^{max} = \Delta S^{max}_{iso}T/C_p$ ($M_{sat}$ is the saturation magnetisation and $C_p$ is the heat capacity).
In our case, the rate of change of $T_N$ with strain is relatively small compared to the large $\Delta S^{max}_{iso}$ in Mn$_3$GaN which would limit $\Delta T_{ad_1}^{max}$ if the elastocaloric-based cooling cycle was restricted to strains below $0.75\%$, as indicated in Fig.~\ref{f_ST}. However, at larger strains the cooling cycle benefits from the additional second-order transition between the collinear magnetic structures and the PM state with high $\partial T_r/ \partial \varepsilon_{xx}$.
This causes a previously unreported qualitative change of the temperature dependence of $S_{tot}(\bm{\varepsilon},T)$. For large enough values of $\varepsilon_{xx}$ the collinear structures are stabilized and two phase transitions are triggered with increasing temperature, namely first-order canted triangular-to-collinear FIM(or AFM) and second-order collinear FIM(or AFM)-to-PM. As a result the adiabatic temperature change is substantially increased from $\Delta T^{max}_{ad_1}$ to $\Delta T^{max}_{ad_2}$  (see Fig.\ \ref{f_ST}).
Hence, our elastocaloric cycle offers simultaneously both large $\Delta S^{max}_{iso} \approx 100$~J/kgK and $\Delta T_{ad_2}^{max} \approx 30$~K in the room temperature range. Even if the corresponding experimental $\Delta S^{max}_{iso}$ was a factor of 5 lower (as suggested by the observed barocaloric effect\cite{matsunami2014giant}) the proposed cycle would still be highly competitive with the available magnetocaloric and mechanocaloric counterparts.\citep{sandeman2012magnetocaloric,moya2014caloric}

We conclude that the combination of the first-order and second-order transitions improves substantially the cooling capacity of the elastocaloric cycle. We stress that both the stability of the collinear magnetic structures and the existence of the first-order transition are underpinned by the strong spin-lattice coupling due to frustrated exchange interactions. In addition, the availability of phase transitions between two ordered states is relevant for elastocaloric-based cooling applications as it can reduce losses due to spin fluctuations and short-range order of a PM state.\cite{pecharsky1997giant}

We note that our cooling cycle relies on large strain-change, $\Delta\varepsilon_{xx} \approx 1\%$. We envisage a device based on a Mn$_3$AN film deposited epitaxially on a piezoelectric perovskite substrate%, adopting the approach of Schleicher et al. aimed at  an electrocaloric bilayer, Ni-Mn-Ga-Co/PMN-PT \cite{schleicher2015epitaxial}. A good elastic coupling between the antiperovskite and perovskite layers is expected. %owing to the matching crystal structure. The lattice mismatch between Mn$_3$GaN and Pb(Mg$_{1/3}$Nb$_{2/3}$)O$_3$-PbTiO$_3$ (PMN-PT) is about~$1.5\%$ and 
such as Pb(Mg$_{1/3}$Nb$_{2/3}$)O$_3$-PbTiO$_3$ which can induce the required strain.\cite{park1997ultrahigh}

\begin{figure}
\includegraphics[width=0.97\columnwidth]{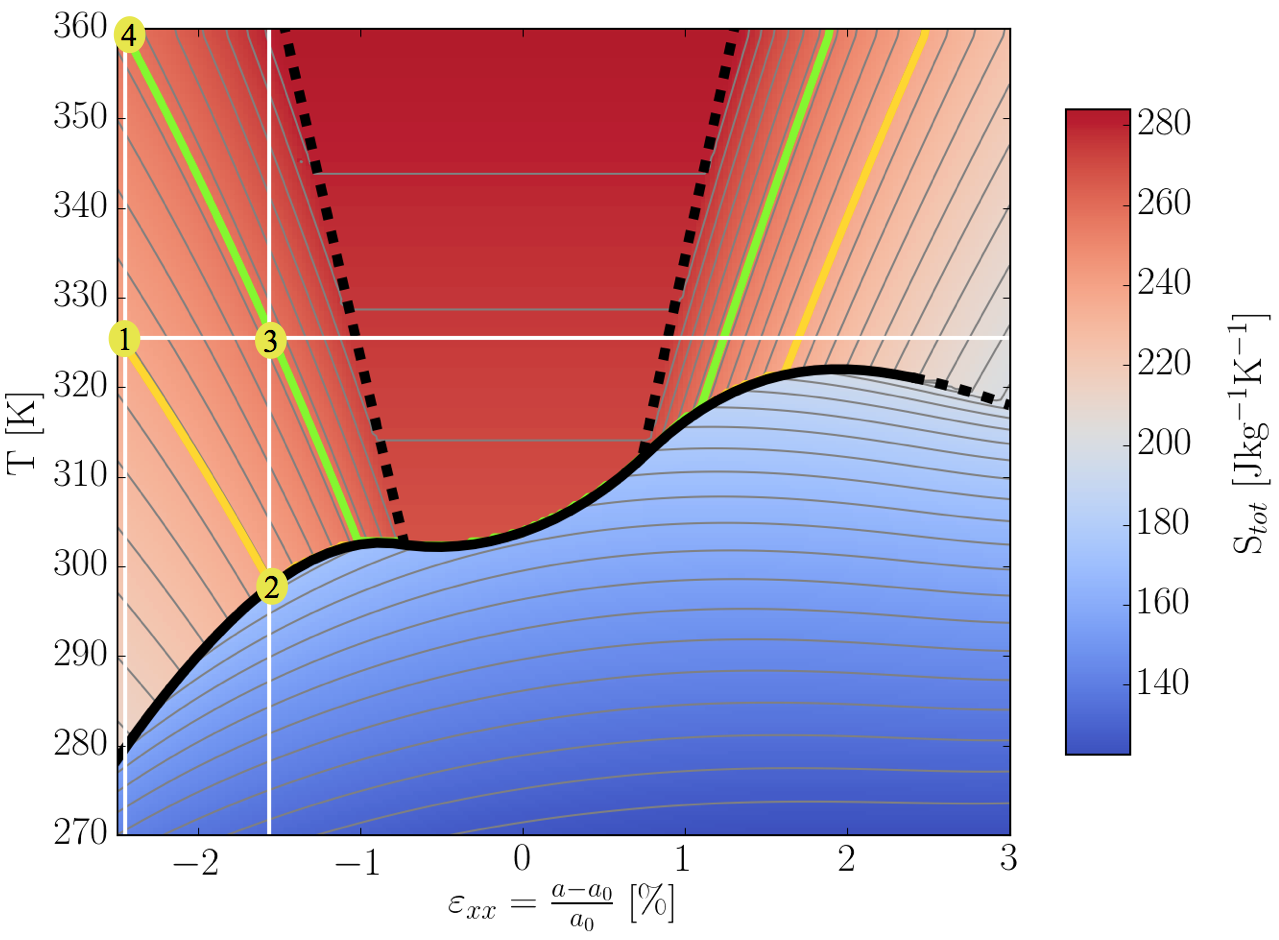}
\caption{(Color online) 2D presentation of the total entropy profile $S_{tot}(\varepsilon_{xx},T)$ of Fig.~\ref{f_eCE}; Black lines mark the first (solid) and second (dashed) order transitions; Numbers mark the stages of an alternative elastocaloric cooling cycle.}
\label{f_eCE_2D}
\end{figure}

\begin{figure}
\includegraphics[width=0.97\columnwidth]{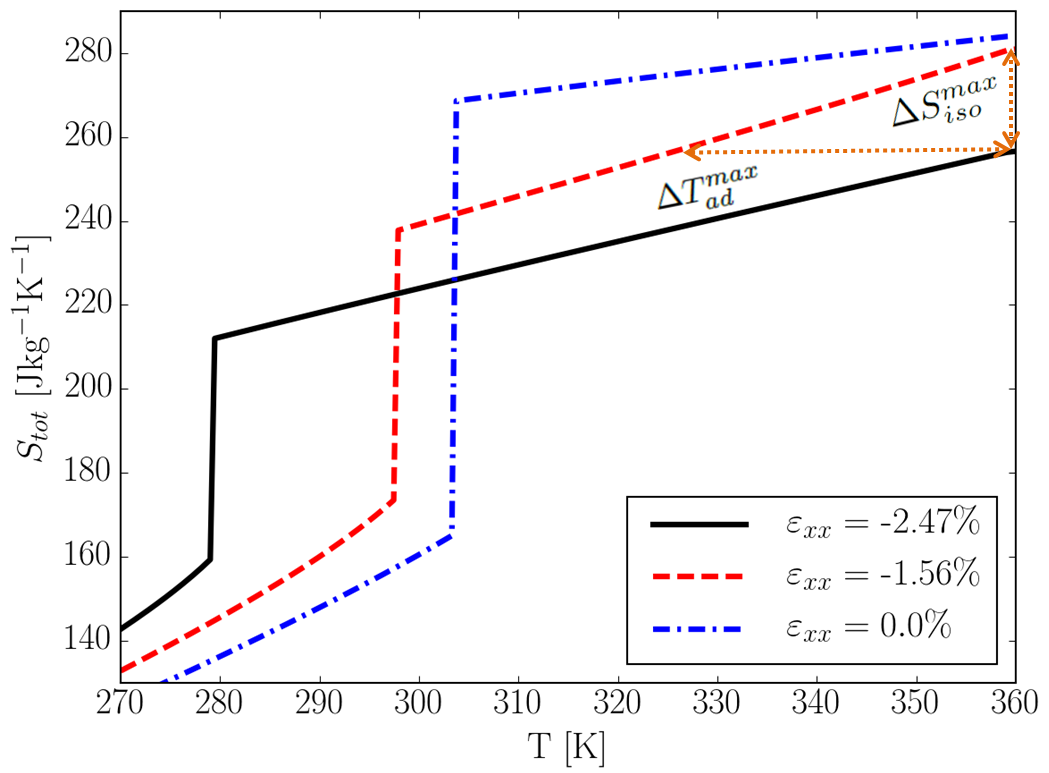}
\caption{(Color online) Total entropy against temperature for three fixed values of strain; The same plot as in Fig.\ \ref{f_ST} of the manuscript but focusing on strains relevant for the alternative cooling cycle. No phase transitions are crossed within the cycle.}
\label{f_SvsT2}
\end{figure}

Finally, we discuss the challenges of losses due to thermomagnetic hysteresis associated with a first-order phase transition. A quantitative analysis would require non-equilibrium thermodynamic modelling\cite{smith2012materials} which is beyond the scope of this work. Experimentally, hysteresis losses have been addressed by tuning phase transitions using field or chemical composition to a crossover between first and second-order behaviour (tricriticallity).\cite{fujita2003itinerant,provenzano2004reduction,trung2009tunable,barcza2010giant,sandeman2012magnetocaloric}
Our phase diagram in Fig.~(\ref{f_PD}) offers a tricritical point at $\varepsilon_{xx} \approx 2.5\%$ (with a reduced entropy change). 
Moreover, the complex entropy profile of Fig.~\ref{f_eCE} allows for construction of cooling cycles not crossing the first-order phase transition. Fig.~\ref{f_eCE_2D} shows such an example which utilizes the strong dependence of entropy on strain in the collinear FIM state. In doing so, it solves the both problem of large magnetothermal hysteresis and the wide required strain span encountered in the cycle of Fig.~\ref{f_eCE}. Furthermore, the equivalent of the maximum adiabatic temperature change (defined for first-order transitions) is not compromised, $\Delta T_{ad}^{max*} \approx 30$~K as shown in Fig.~\ref{f_SvsT2}. However, the maximum entropy change comes down to $\Delta S^{max*}_{iso} \approx 20$~J/kgK as expected for a continuous change of magnetic ordering. 
Alternatively, a "multicaloric" cycle using a combination of strain and magnetic field could harness the sizeable strain-induced moment in Mn$_3$AN to facilitate a transfer of hysteresis losses between magnetic and elastic energy, following a recent example of FeRh.\cite{liu2016large}
We hope that these predictions will motivate further experimental study.

\section{Conclusions}

In summary we have modelled the geometrically frustrated magnetic structure of non-collinear and collinear magnetic structures in Mn-antiperovskite nitrides with relaxed and biaxially strained lattice at zero and at finite temperatures. Firstly, by performing extensive SDFT simulations at zero temperature, we have systematically investigated the piezomagnetic effect. Remarkably, we have linked the sign of the canted angle  to a simple feature of the band structure: the relative energy separation between the $d$-states of atom A and Mn.

Secondly, we have developed a SDFT-based disordered local moment theory to study the impact of finite temperature on the magnetic ordering and to evaluate the elastocaloric effect.  We have applied the theory to relaxed Mn$_3$GaN and found the stability of the triangular AFM phase at low temperature and a first-order transition to PM phase at the N\'eel temperature all in good agreement with available experimental data. The theory is able to provide the relevant thermodynamic quantities and to describe the stability of competing magnetic phases which allowed us to construct a strain-temperature magnetic phase diagram. We predict two novel magnetic phases, namely collinear ferrimagnetic at $\varepsilon_{xx}<$ 0.75\% and collinear antiferromagnetic at $\varepsilon_{xx}>$ 0.75\%. These collinear structures are stable at high temperatures and show a second-order transition to the PM state which strongly depends on biaxial strain. The combination of both second-order and first-order transitions enabled us to propose an elastocaloric cooling cycle which exhibits large isothermal entropy change and adiabatic temperature change simultaneously. This rich phenomenology is available due to the strong spin-lattice coupling linked fundamentally to the magnetic frustration.

We conclude that the Mn$_3$AN family of frustrated non-collinear AFMs with complex phase diagrams represent ample opportunity to tune the chemical composition and control the critical stimuli to achieve similar or better cooling characteristics than shown here while still utilizing relatively abundant chemical elements. We thus suggest Mn$_3$AN as a new class of elastocaloric materials.

%%%%%%%%%%%%%%%%%%%%%%%%%%%%%%%%%%%%%%%%%%%%
\begin{acknowledgments}
We would like to thank Lesley F. Cohen, William R. Branford, Andrei Mihai, Bin Zou, David C. Boldrin, and Christopher E. Patrick, for productive discussions.
The research has received funding from the European Community’s 7th Framework Programme under Grant agreement 310748 “DRREAM”. The work at the University of Warwick was supported by the U.K.~EPSRC, grants EP/J06750/1 and EP/M028941/1.
\end{acknowledgments}

%%%%%%%%%%%%%%%%%%%%%%%%%%%%%%%%%%%%%%%%%%%%
%Appendix
%%%%%%%%%%%%%%%%%%%%%%%%%%%%%%%%%%%%%%%%%%%%

\appendix

\section{Supercell-based \textit{ab initio} calculation}

\begin{figure}
\includegraphics[width=0.9\columnwidth]{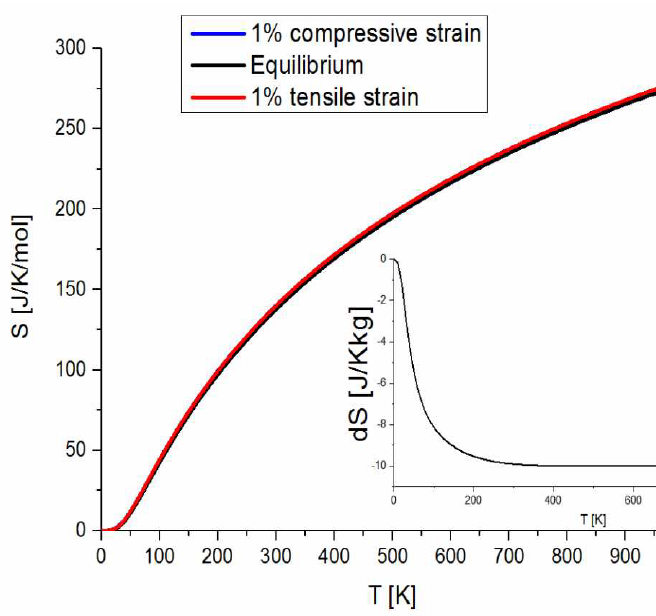}
\caption{(Color online) Lattice vibrational entropy of Mn$_3$GaN with canted triangular magnetic order vs temperature for the cubic and two tetragonal cases; The inset shows $S_{ph}(\varepsilon_{xx}=0,T)-S_{ph}(\varepsilon_{xx}=\pm 1\%,T)$.}
\label{f_Sph}
\end{figure}

Performing supercell-based $ab~initio$ phonon calculations for the full range of required temperatures and strains would be computationally too demanding. 
To gauge the model size of the lattice entropy change and the adiabatic temperature change of the Debye lattice vibrations we performed only one such calculation for canted triangular state to compare the lattice entropy of the cubic, -1\% compressive, and 1\% tensile-strained system using VASP and Phonopy.\cite{togo2015first} We used a 2x2x2 superlattice formed of magnetic unit cells with 5 atoms (40 atoms in total). Calculations for forces on atoms were performed in the non-collinear regime including spin-orbit coupling. We obtained the phonon dispersion relations and densities of states using Phonopy (finite displacement method). The resulting $S_{ph}$ is plotted in Fig.~\ref{f_Sph} as a function of temperature.
We obtain an isothermal entropy change $S_{ph}(\epsilon_{xx}=-1\%) - S_{ph}(\epsilon_{xx}=0) \approx S_{ph}(\epsilon_{xx}=1\%) - S_{ph}(\epsilon_{xx}=0) \approx 10$~J/kgK at 300~K and above. Although this value is large, it is relatively small compared to the $\Delta S^{max}_{mag}$=103.2 J/kgK obtained at the first-order transition at zero strain. We therefore conclude that $S_{tot}$ is dominated by $\bar{S}_{mag}$.

\section{DLM linear response theory}\label{DLMlinresp}
Our approach is based on magnetic susceptibility which quantifies the response of the magnetic system to an infinitesimally small site-dependent magnetic field. In the PM state we can make use of the high symmetry of the system to write:
\begin{equation}
\sum_{j} \left[3k_B T \delta_{i,j} - S^{(2)}_{i,j}(\textbf{q})\right] \chi_{j,k}(\textbf{q},T) = \mu^2 \delta_{i,k}
\end{equation}
where $S^{(2)}_{i,j}(\mathbf{q})$ is the lattice Fourier transform of the spin-spin correlation function in direct space, $\mu$ is the size of the local moments in the fully disordered state, and the indices $i$, $j$, $k$ run through the sites with local magnetic moments. The eigenvectors of matrix $[3k_B T \delta_{i,j} - S^{(2)}_{i,j}(\textbf{q})]$ and the $\textbf{q}$-vector for the largest eigenvalue give full information about the favoured magnetic instability.\cite{gyorffy1985first,staunton1986static} 
We use the KKR multiple scattering theory to evaluate $S^{(2)}_{i,j}(\textbf{q})$ without assuming a particular type or periodicity of the antiferromagnetic order.

\section{Computational details of SDFT-based DLM theory}

\begin{figure}
\includegraphics[width=0.85\columnwidth]{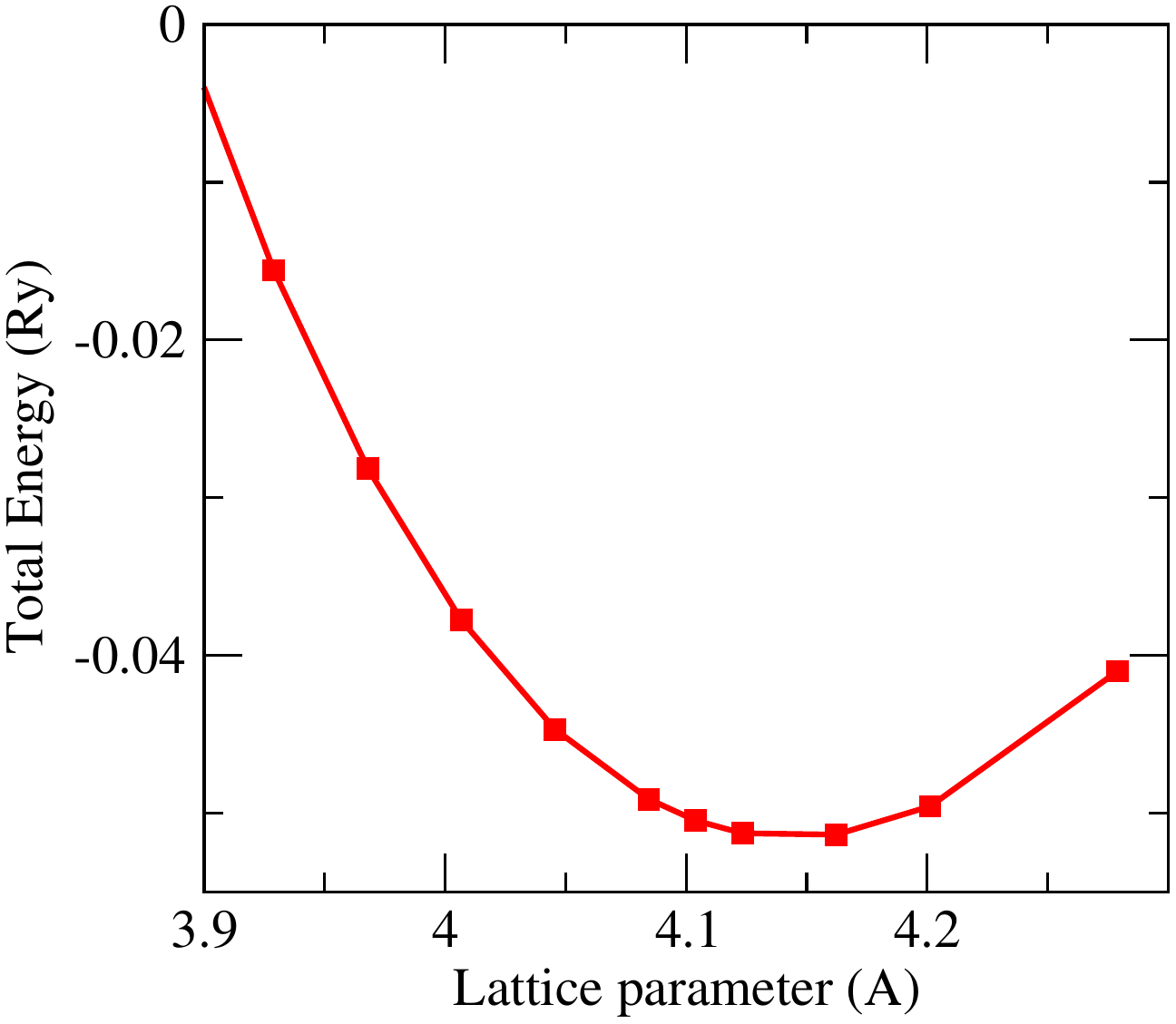}
\caption{(Color online) Total energy calculated at $T=0K$ from the SDFT code against different values of the lattice parameter for the cubic lattice in Mn$_3$GaN. The energies are given with respect to the total energy we obtain for the experimental lattice system.}
\label{TotalE}
\end{figure}

We used a multiple scattering Korringa-Kohn-Rostoker (KKR) theory based on green's function formalism.\cite{staunton2006temperature,stocks1978complete,johnson1986density,ebert2011calculating} As mentioned in the manuscript the average over the local moment orientations was carried out via the coherent potential approximation.\cite{stocks1978complete,ebert2011calculating}
For the treatment of the SDFT potentials we used the muffin-tin approximation and the maximum value of the angular momentum used to describe the scattering of the radial problem was set to $l_{max}$=3.
The Weiss field SDFT-DLM data generated to fit the expansion coefficients of the internal magnetic energy was obtained from the self-consistent calculation of charge and magnetization densities for the paramagnetic state.
In case of Eq.~(\ref{FreeEcub}) there is only one order parameter $m$ and two coefficients. We found that seventeen different values of the quantity  $\beta\bm{h}=-\beta\nabla_m\bar{\Omega}$ (describing the triangular state), ranging from 0.05 to 10, were enough to fit the constants $a$ and $b$. These values correspond to the order parameter $m$ varying from 0.02 to 0.9. In case of the more complicated Eq.~(\ref{FreeEtet}) the ten expansion coefficients were extracted from more than two hundred independent calculations of the quantities $\{\beta\bm{h}_i\}$, comprising different triangular AFM, distorted triangular AFM, and collinear FIM and AFM magnetic structures. The error for both fits was within $\approx 1\%$.

Prior to exploring the strained systems, we identified the lattice parameter of the cubic unit cell that minimizes our KKR-based total energy. Fig.~\ref{TotalE} shows the total energy against the lattice parameter of the cubic lattice in Mn$_3$GaN at zero temperature. The total energy minimizes at 4.14$\buildrel _{\circ} \over {\mathrm{A}}$, which is roughly 6$\%$ higher compared to the experimental value of  3.898$\buildrel _{\circ} \over {\mathrm{A}}$.\cite{takenaka2014magnetovolume}
The magnetic moments $\mu\approx 3.1\mu_\text{B}$ localised at each Mn site in the unstrained system are slightly larger than the corresponding value $\mu\approx 2.43\mu_\text{B}$ obtained by VASP with the equilibrium lattice parameter of 3.86$\buildrel _{\circ} \over {\mathrm{A}}$. The fast increase of the local moment size with increasing unit cell volume is another consequence of the geometric frustration of the exchange interactions.

\begin{figure}
\includegraphics[width=0.85\columnwidth]{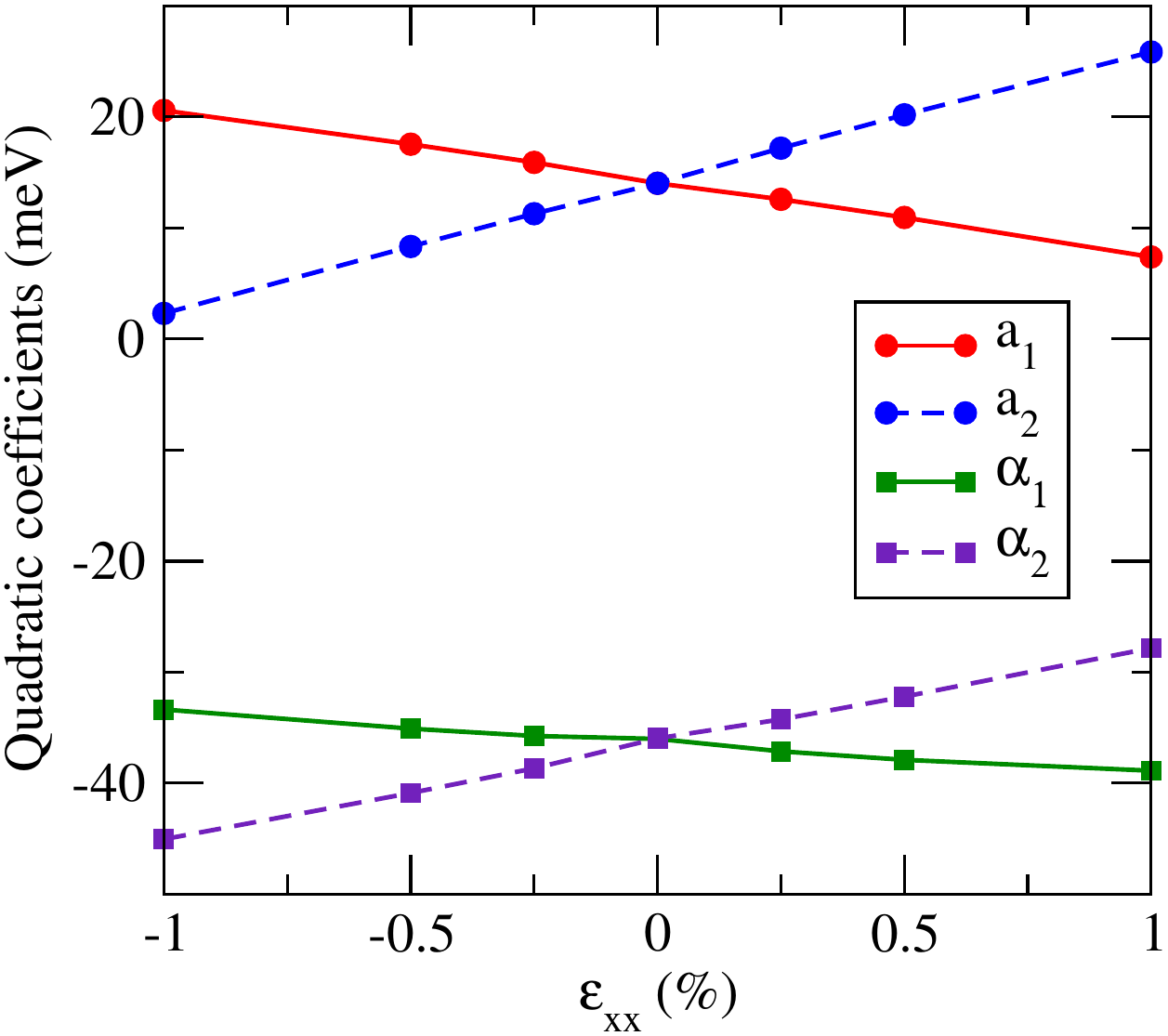}
\caption{(Color online) The second order coefficients $a_1$, $a_2$, $\alpha_1$, and $\alpha_2$ obtained from fitting the SDFT-DLM Weiss field data for $\varepsilon_{xx}$=0\%, $\pm$0.25\%, $\pm$0.5\%, $\pm$1.0\%. The change of the quartic coefficients has been found to have no relevant effect on our calculations.}
\label{2nd}
\end{figure}

%It should be noted that the calculation starts with obtaining the charge and magnetization potentials self-consistently for the PM state which maps to an Ising picture with one half of the moments oriented antiparallel the to other half, on average (all order parameters equal to zero). This symmetry is then broken in a magnetically ordered state and the ensemble averages over the non-collinear local moment configurations are needed to predict the observable magnetic properties (finite order parameters).

Fig.~\ref{2nd} shows the linear dependence of the quadratic coefficients $\{a_1, a_2, \alpha_1, \alpha_2\}$ on biaxial strain $\varepsilon_{xx}$. The change of the quartic coefficients with $\varepsilon_{xx}$ was found to have negligible effect on the temperature-strain magnetic phase diagram. Their values at zero strain are $b_1$=$b_2$=2.856 meV, $\beta_1$=$\beta_3$=-37.54 meV, $\beta_2$=$\beta_4$=43.08 meV.

Finally, at low temperatures our DLM simulations predict canting and change of size of local moments of the triangular phase in Fig.~\ref{f_structure}~(c,d) in good agreement with our zero-temperature results\cite{zemen2015piezomagnetic} underlying the PME data presented in Fig.~\ref{f_pme}. 
In the case of Mn$_3$GaN we found $\partial \theta_1/\partial \varepsilon_{xx} \approx -4$~deg at 5K, which is in  semiquantitative agreement with our VASP simulations, $\theta_1/\partial \varepsilon_{xx} \approx -2.2$~deg.
%The discrepancy could be due to the difference in equilibrium lattice parameters at zero temperature obtained by the VASP code and DLM-DFT (see appendix B).


\end{document}